\documentclass[twocolumn,preprintnumbers,elsart]{revtex4}
\usepackage{eurosym}
\usepackage{makeidx}
\usepackage{amssymb}
\usepackage{amsmath}
\usepackage{mathrsfs}
\usepackage{graphicx}
\usepackage{dcolumn}
\usepackage{bm}
\usepackage[center]{subfigure}
\usepackage{color}
\usepackage{indentfirst}
\usepackage{array}
\usepackage{booktabs}
\usepackage{multirow}
\usepackage{upgreek}

\begin{document}

\title{Can vortex quantum droplets be realized experimentally?}
\author{Guilong Li$^{1}$}
\thanks{These authors contributed equally to this work.}
\author{Zibin Zhao$^{1}$}
\thanks{These authors contributed equally to this work.}
\author{Bin Liu$^{1,2}$}
\email{binliu@fosu.edu.cn}
\author{Yongyao Li$^{1,2}$}
\email{yongyaoli@gmail.com}
\author{Yaroslav V. Kartashov$^{3}$}
\author{Boris A. Malomed$^{4,5}$}
\affiliation{
$^{1}$School of Physics and Optoelectronic Engineering, Foshan University,
Foshan 528225, China\\
$^{2}$Guangdong-Hong Kong-Macao Joint Laboratory for Intelligent Micro-Nano
Optoelectronic Technology, Foshan University, Foshan 528225, China\\
$^{3}$Institute of Spectroscopy, Russian Academy of Sciences, 108840 Troitsk, Moscow, Russia\\
$^{4}$Department of Physical Electronics, School of Electrical Engineering,
Faculty of Engineering, Tel Aviv University, Tel Aviv 69978, Israel\\
$^{5}$Instituto de Alta Investigaci\'{o}n, Universidad de Tarapac\'{a},
Casilla 7D, Arica, Chile}

\begin{abstract}
The current state of research on vortices carried by quantum droplets (QDs) has predicted their existence, in the stable form, in two- and three-dimensional free-space binary Bose-Einstein condensates (BECs) and dipolar BECs. These theoretical results suggest that QDs may be excellent carriers of self-trapped vortex states. Given that the experimental creation of QDs has already been firmly established, the observation of embedded vortices in them becomes a key question for the next phase of the development in the field.
\end{abstract}

\maketitle



Creation of stable self-trapped vortex states (alias vortex solitons) in
Bose-Einstein condensates (BECs) in free space is a challenging problem \cite%
{Malomed2022}. To date, an experimental demonstration of this possibility is
missing. A new powerful platform for the realization of various self-trapped
states is offered by quantum droplets (QDs), i.e., stable localized modes
maintained by the balance of mean-field (MF) interactions and corrections to
them induced by quantum fluctuations \cite{LHY}. Recent theoretical
predictions \cite{Petrov2015,Petrov2016} and experimental demonstrations
\cite%
{Ferrier-Barbut2016,Schmitt2016,Cabrera2018,Cabrera2,Inguscio2018,DErrico2019}
suggest various possibilities for the creation of novel fundamental and
vortical patterns, as summarized in reviews \cite%
{Luo2021,Guo2021,Bttcher2021}. In particular, QDs with embedded vorticity
were theoretically investigated in detail \cite%
{Kartashov2018,Cidrim2018,Li2018,Li2024FOP,Li2024PRL,Dong2023,Dong2024,Salgueiro2024}.

This story starts with the original experimental realization of BECs \cite%
{Anderson1995,Davis1995,Bradley1995}, which has made it possible to probe
properties of quantum matter that are otherwise difficult to access. By
observing macroscopic quantum phenomena manifested by BECs, researchers gain
valuable insights into the underlying physical phenomena and mechanisms \cite%
{Pethick2002,Pitaevskii2016,Yongchang2019,Yongchang2021,Xiaoling2021,Guo2019,He2024,Pan2022,Liu2023,Gu2023,Deng2023,Peng2023}.

In experimental settings, the BEC lifetime is usually short, as the
condensates are prone to spatial expansion or collapse, in the cases of the
repulsive and attractive inter-atomic interactions, respectively \cite%
{Donley2001}. The lifetime may be substantially extended by means of a
trapping potential which confines the condensate \cite%
{Pethick2002,Pitaevskii2016,Malomed2022}.

It is relevant to stress that the natural interaction between atoms is
repulsive, due to the fact that colliding hard particles bounce back from
each other. The interaction may be switched to attraction by dint of the
Feshbach-resonance technique, implemented by dc magnetic field or uniform
laser illumination applied to BEC \cite{Chin2010,Courteille1998}. However,
in the multidimensional geometry the attractive interaction readily leads to
the critical or supercritical collapse (alias blowup), in the two- and
three-dimensional (2D and 3D) cases, respectively \cite%
{Berge,Fibich,Malomed2022}. In terms of theoretical modeling, the respective
Gross-Pitaevskii equations (GPEs) \cite{Pethick2002,Pitaevskii2016} produce
a true singularity after a finite evolution time, while in the experiment
the collapse leads to destruction of the BEC state. The collapse tends to
destabilize the multidimensional solitons, unlike their 1D counterparts,
which are usually stable \cite{Zakharov1980} (in BEC, the stability of the
effectively 1D matter-wave solitons has been demonstrated in well-known
experiments \cite{Hulet,Khaykovich2002,Wieman}).

Therefore, the creation of physically relevant self-trapped 2D and 3D states
in BECs, which should be stable against the expansion and collapse, is a
challenging topic, with important implications not only for BEC but also
with interesting analogies in the fields of nonlinear optics and photonics,
plasma physics, etc. \cite{Malomed2022}. In this context, BECs offer an
ideal platform for the prediction and observation of the dynamics and
stability of nonlinear self-trapped states in the multidimensional space. In
particular, the multidimensional space offers possibilities for building
states with integer intrinsic vorticity (alias the topological charge or
winding number) \cite{Malomed2019}. However, vortex self-trapped states
(vortex solitons) are subject to azimuthal modulational instability which
leads to spontaneous splitting of vortices into separating fragments, prior
to the onset of the collapse, as the splitting instability is stronger than
its collapse-driven counterpart (later, the fragments may suffer the
intrinsic blowup) \cite{Malomed2022,Malomed2019,Malomed2002,Torner1997,Firth1997,Torres1998,Kruglov1985,Bigelow2004,Petrov1998,Minardi2001,Tikhonenko1995}.

Under the action of specific potential traps, quantized vortices were
experimentally observed in BEC with the self-repulsive nonlinearity \cite%
{Henn2024,Madison1999,Madison2000,Ku2014,Donadello2014,Klaus2022,Casotti2024}%
, which is obviously impossible in the free space. On the other hand, the
interplay of a spatially periodic optical-lattice potential with the MF
cubic self-repulsion was predicted to support stable 2D gap solitons with
embedded vorticity \cite{HS,Ostrovskaya2004}.

Promising theoretical predictions of matter-wave solitons with embedded
vorticity were produced making use of the linear spin-orbit coupling in
binary BEC. As a result, stable 2D \cite{Ben-Li,Jiang2016,Deng} and
metastable 3D \cite{Han-Pu} solitons have been predicted, in the form of
semi-vortices, as only one component carries the vorticity in them. Also
predicted were stable vortex solitons in ultracold gases of Rydberg atoms
\cite{Rydberg,Rydberg2}, and in microwave-coupled two-component BEC \cite%
{Qin}.

The theoretical and experimental studies of multidimensional self-trapped
objects (both fundamental (zero-vorticity) and vortex solitons) have been
significantly propelled by the above-mentioned QD concept, realized in
two-component (binary) BEC in the three-dimensional geometry, with the cubic
MF attraction between the components slightly dominating over the MF
intra-component self-repulsion. The overall stabilization is provided by the
higher-order (quartic) beyond-MF (BMF)\ repulsion. The balance between the
residual MF attraction and BMF repulsion gives rise to QDs filled by the
ultradilute superfluid \cite{Petrov2015}. The QDs were observed
experimentally in monoatomic binary BECs realized in the ultracold gas of $%
^{39}$K atoms \cite{Cabrera2018,Cabrera2,Inguscio2018}, and heteroatomic $%
^{41}$K--$^{87}$Rb \cite{DErrico2019} and $^{23}$Na--$^{87}$Rb \cite{Guo2022}
mixtures. The balance corresponds to a minimum of the total energy of the
binary BEC for a given number of particles in the condensate \cite%
{Ferrier-Barbut2018}. In addition to the scheme involving binary
condensates, single-component BECs of magnetic atoms can also form dipolar
QDs, as demonstrated experimentally in the BEC of $^{164}$Dy \cite%
{Ferrier-Barbut2016,Schmitt2016}, $^{166}$Er \cite{Chomaz2016}, and $^{162}$%
Dy \cite{Bttcher2019} atoms (in the absence of the BMF stabilization,
dipole-dipole interactions between magnetic atoms in the 3D free space give
rise to the $d$-wave collapse \cite{Lahaye2008}).

The above-mentioned achievements in the experimental creation of stable
quasi-2D \cite{Cabrera2018,Cabrera2} and 3D \cite{Inguscio2018} QDs strongly
suggest to investigate the possibility of predicting stably self-trapped
two-component QDs with embedded vorticity, in the 2D \cite{Li2018} and 3D
\cite{Kartashov2018} free space (on the contrary to these results,
single-component vortex QDs in the 2D dipolar BEC with the simplest
isotropic structure are unstable \cite{Cidrim2018}). Also predicted were
stable vortex QDs with \textquotedblleft exotic" anisotropic shapes \cite%
{Li2024FOP,Li2024PRL,Kartashov2020}.

Here we aim to briefly review the theoretical results for the QDs with
embedded vorticity (a somewhat broader review on a related topic was very
recently published in Ref. \cite{Chen2024}). First, on the basis of the BMF amendment to the MF
theory, which arises from the Bogoliubov excitations around the MF state, as
elaborated long ago by Lee, Huang, and Yang \cite{LHY}, the corresponding
correction to the BEC energy density was derived by Petrov \cite{Petrov2015}%
:
\begin{equation}
E_{\mathrm{BMF}}=\frac{128}{30\sqrt{\pi }}gn^{2}\sqrt{na_{s}^{3}},
\label{BMF}
\end{equation}%
where $a_{s}$ is the $s$-wave scattering length, $n$ are equal density of
the two components of the binary BEC, $g=4\pi \hbar ^{2}m/a_{s}$ represents
the strength of the contact interaction, and $m$ is the atomic mass.

\begin{figure*}[htbp]
{\includegraphics[width=2\columnwidth]{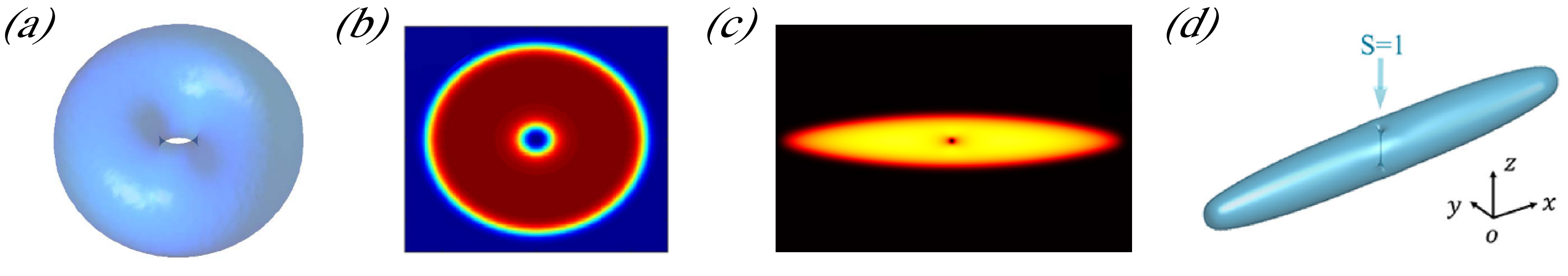}}
\caption{Typical examples of predicted stable vortex QDs with topological
charge $S=1$. (a) In the 3D binary BECs, as per Ref. \protect\cite%
{Kartashov2018}. (b) In the 2D binary BECs, as per Ref. \protect\cite{Li2018}
. (c) In the 2D dipolar BEC, as per Ref. \protect\cite{Li2024FOP}. (d) In
the 3D dipolar BEC, as per Ref. \protect\cite{Li2024PRL}.}
\label{example}
\end{figure*}

Using the systems of LHY-amended GPEs for the binary condensate, 3D
vortex-QD solutions with the toroidal structure, carrying the topological
charge $S=1$ and $2$ were constructed in Ref. \cite{Kartashov2018}. An
example of the stable 3D self-trapped vortex torus with $S=1$ is displayed
in Fig. \ref{example}(a).

Further, the dimensional reduction 3D $\rightarrow $ 2D, imposed by strong
confinement (trapping potential) acting in the third direction, produces the
BMF correction to the 2D energy density in the form of \cite{Petrov2016}
\begin{equation}
E_{\mathrm{2DBMF}}=\frac{8\pi n^{2}}{\ln ^{2}(\left\vert a_{\uparrow
\downarrow }\right\vert /a)}[\ln (n/(n_{0})_{\mathrm{2D}})-1],  \label{2DBMF}
\end{equation}%
where $n_{\upuparrows }=n_{\downdownarrows }\equiv n$ are equal densities of
the \textquotedblleft top" and \textquotedblleft bottom" BEC\ components (if
the MF\ wave function is construed as a spinor), $a_{\upuparrows
}=a_{\downdownarrows }\equiv a>0$ are equal scattering lengths of the
repulsive intra-component interactions, $a_{\uparrow \downarrow }<0$ is the
negative scattering accounting for the inter-component attraction, and $%
(n_{0})_{\mathrm{2D}}$ is the equilibrium density. The change of the sign of
expression (\ref{2DBMF}) from negative to positive with the increase of $n$
implies the arrest of the critical collapse, and it was demonstrated in Ref.
\cite{Li2018} that the corresponding BMF-amended 2D GPE produced vortex-QD
solutions which may be stabilized not only against the collapse, but also
against the azimuthal modulational instability (splitting), see an example
of a stable vortex mode with $S=1$ in Fig. \ref{example}(b). Due to the
presence of the logarithmic factor in Eq. (\ref{2DBMF}), these QDs tend to
exhibit a flat-top shape, expanding with the increase of the norm and
vorticity. The results demonstrate that such vortex QDs keep a stability
area in the respective parameter area up to $S=5$, which is unusual for
vortex solitons in nonlinear media. It should be mentioned that in addition to vortex QDs condensates with LHY quantum correction may support rich variety of 2D or 3D dynamical cluster-like states with nontrivial phase distributions that exhibit rotation and pulsations in the course of evolution \cite{Kartashov2019,Dong2021,Tengstrand2019}.

The use of dipolar BECs in ultracold gases of magnetic atoms opens new
perspective for the creation of stable vortex QDs. The energy of the
long-range dipole-dipole interaction (DDI) for polarized magnetic moments $%
\mu $ is given by the usual expression \cite{Lahaye2009,Stuhler2005}:
\begin{equation}
V_{\mathrm{DDI}}=\frac{\mu _{0}\mu ^{2}}{4\pi }\frac{1-3\cos ^{2}{\theta }}{%
r^{2}},  \label{DDI}
\end{equation}%
where $\theta $ is the angle between the vector of length $r$ connecting the
dipoles and the polarization direction, $\mu _{0}$ being the vacuum
permeability. This interaction potential demonstrates the anisotropic nature
of DDIs, paving the way for the emergence of novel phenomena and anisotropic
structures.

As mentioned above, the simplest isotropic (cylindrically symmetric) 3D
vortex-QD states in the dipolar BEC, with the vorticity axis aligned with
the polarization direction of the atomic magnetic moments (imposed by
externally applied magnetic field) and winding number $S=1$, are unstable
against splitting or the onset of vortex-line corrugation caused by
Kelvin-wave excitations \cite{Cidrim2018}. Anisotropic QDs with the
vorticity axis directed perpendicular to the polarization of the magnetic
moments were investigated in Refs. \cite{Li2024FOP,Li2024PRL}. This
configuration transforms the hollow core of the vortex from a state
associated with lower negative attraction energy to one that removes some
positive repulsion energy, resulting in a lower overall energy and,
eventually, \emph{stability} of the resulting vortex QDs with the
anisotropic shape.

In Ref. \cite{Li2024FOP} it was proposed, leveraging the established
experimental technique \cite%
{Burger1999,Leanhardt2002,Andrelczyk2001,Scherpelz2014,Shen2019}, to create
2D anisotropic vortex QDs by phase-imprinting a dark soliton onto an
original zero-vorticity dipolar QD, allowing it to transform into a vortex.
Subsequently, the analysis was extended to the 3D free space, predicting
\emph{stable} strongly anisotropic vortex QDs \cite{Li2024PRL}, see an
example in Figs. \ref{example}(c,d). Even in the presence of loss effects
induced by three-body collisions, the 3D vortex persists in the course of
long evolution. The robustness of the so predicted anisotropic vortex QDs is
highly encouraging for their creation in the experiment.

To date, a clear picture has emerged: the persistent challenge of seeking
for stable self-trapped vortices in BECs may be resolved using the powerful
QD platform. The above-mentioned demonstrations of the fundamental
(zero-vorticity) QDs in various experimental setups and theoretical
predictions of the stable QDs with embedded stability provide a strong
incentive for the further work in this direction. A possible experimental
approach may rely on injecting the topological charge into fundamental QDs,
using available technologies such as phase imprinting \cite%
{Leanhardt2002,Andrelczyk2001,Scherpelz2014,Shen2019} and magnetosrirring
\cite{Klaus2022,Casotti2024}. If the prediction of the stable vortex QDs is
experimentally validated, it will imply the first realization of
self-trapped vortices in the free-space BEC.

\textit{Acknowledgments}. This work was supported by NNSFC (China)
through Grants No. 12274077, No. 12475014, No. 11874112 and No. 11905032, the Natural Science Foundation of Guangdong province through
Grant No. 2024A1515030131, No. 2023A1515010770, the Research Fund of Guangdong-Hong
Kong-Macao Joint Laboratory for Intelligent Micro-Nano Optoelectronic
Technology through grant No.2020B1212030010. The work of B.A.M. was supported, in part, by the
Israel Science Foundation through grant No. 1695/2022. Y.V.K. acknowledges
funding by research project FFUU-2024-0003 of the Institute of Spectroscopy
of the Russian Academy of Sciences.


\end{document}